\documentclass[8.5pt,onecolumn]{article}
\usepackage[super,sort&compress,comma]{natbib}
\usepackage{mhchem}
\usepackage{times}
\usepackage{balance}

\usepackage{graphicx} 
\usepackage[format=plain,justification=raggedright,singlelinecheck=false,font=small,labelfont=bf,labelsep=space]{caption}
\usepackage{fancyhdr}
\usepackage{caption}

\usepackage{rotating}
\usepackage{color}

\fancypagestyle{plain}{
}

\makeatletter
\def\subsubsection{\@startsection{subsubsection}{3}{10pt}{-1.25ex plus -1ex minus -.1ex}{1ex plus 1ex}{\normalsize\bf}}
\def\paragraph{\@startsection{paragraph}{4}{10pt}{-1.25ex plus -1ex minus -.1ex}{0ex plus 0ex}{\normalsize\textit}}
\renewcommand\@biblabel[1]{#1}
\renewcommand\@makefntext[1]%
{\noindent\makebox[0pt][r]{\@thefnmark\,}#1}
\makeatother

\usepackage{amsmath}
\usepackage{amssymb}
\usepackage{amsmath,amssymb}
\usepackage[english]{babel}
\usepackage{bm}
\usepackage{color}
\usepackage{indentfirst}

\newcommand{\angstrom}{\mbox{\normalfont\AA}}

\usepackage[normalem]{ulem}
\usepackage{color}
\definecolor{alertcol}{rgb}{0.7,0, 0}
\definecolor{ascol}{rgb}{0.7,0, 0}
\definecolor{yscol}{rgb}{0, 0.4, 0.0}

\begin{document}


\noindent\LARGE{\textbf{Automated Calculation of Thermal Rate Coefficients using Ring Polymer Molecular Dynamics and Machine-Learning Interatomic Potentials with Active Learning}}

\vspace{0.6cm}

\noindent\large{\textbf{I. S. Novikov \textit{$^{a}$}},
\large{\textbf{Y. V. Suleimanov\textit{$^{b,c\ast}$}}} and \large{\textbf{A. V. Shapeev\textit{$^{a\ast}$}}}\vspace{0.6cm}

\date{\today}

\noindent \normalsize{We propose a methodology for fully automated calculation of thermal rate coefficients of gas phase chemical reactions, which is based on combining the ring polymer molecular dynamics (RPMD) with the machine-learning interatomic potentials actively learning on-the-fly. Based on the original computational procedure implemented in the RPMDrate code, our methodology gradually and automatically constructs the potential energy surfaces (PESs) from scratch with the data set points being selected and accumulated during the RPMDrate simulation. Such an approach ensures that our final machine-learning model provides reliable description of the PES which avoids artifacts during exploration of the phase space by RPMD trajectories. We tested our methodology on two representative thermally activated chemical reactions studied recently by RPMDrate at temperatures within the interval of 300--1000~K. The corresponding PESs were generated by fitting to only a few thousands automatically generated structures (less than 5000) while the RPMD rate coefficients retained the deviation from the reference values within the typical convergence error of RPMDrate. In future, we plan to apply our methodology to chemical reactions which proceed via complex-formation thus providing a completely general tool for calculating RPMD thermal rate coefficients for any polyatomic gas phase chemical reaction.} \vspace{0.5cm}

\footnotetext{\textit{$^{a}$~Skolkovo Institute of Science and Technology, 
Skolkovo Innovation Center, Nobel St. 3, Moscow 143026, Russia}}
\footnotetext{\textit{$^{b}$~Computation-based Science and
Technology Research Center, Cyprus Institute, 20 Kavafi Street,
Nicosia 2121, Cyprus}} 
\footnotetext{\textit{$^{c}$~Department of
Chemical Engineering, Massachusetts Institute of Technology,
Cambridge, Massachusetts 02139, United States}}

\footnotetext{$^{\ast}$~Email:ysuleyma@mit.edu, a.shapeev@skoltech.ru}

\section{Introduction}
Accurate and efficient computation of thermal chemical reaction rate coefficients represents one of the most challenging problems for modern theoretical physical chemistry. Rigorous quantum dynamics calculations scale exponentially with the dimensionality of the system and are therefore limited to chemical reactions involving just a few atoms.~\cite{QDreview} Classical description of chemical reactivity allows practical simulations of polyatomic systems, but the problem is complicated at low temperatures, at which quantum-mechanical effects of nuclear motions such as zero-point energy, tunneling, and resonance effects become critically important (though the contribution of the later effect to thermal rate coefficients is less studied~\cite{hguo_review}). Recently, it has been demonstrated that the ring polymer molecular dynamics (RPMD) ~\cite{Craig-Manolopoulos:04,RPMDREVIEW} provides systematically accurate approach for calculating thermal rate coefficients in multifarious scenarios.~\cite{Suleimanov-etal:16} This semiclassical method scales ``classically'' with the number of atoms and is based on the isomorphism between the quantum statistical mechanics of the physical system and the classical statistical mechanics of a fictitious ring polymer consisting of $n_{\rm beads}$ copies (beads) of the original system connected by harmonic springs.~\cite{Wolynes} RPMD is exact in the high-temperature limit as it converges to classical molecular dynamics. It has been also shown that RPMD rate-theory gives a lower bound to RPMD transition state theory, which describes an instantaneous quantum flux from reactants to products~\cite{hele2013-QTSTuniqueness} and describes fluctuations around the instanton geometry (in the deep-tunnelling regime)~\cite{richardson2009-RPMDdeeptunneling}, thus explaining why RPMD provides reliable estimates of the quantum rate coefficient at low temperatures.
General computational procedure for calculating RPMD rate coefficients for polyatomic chemical reactions has been developed~\cite{Suleimanov-etal:11} and implemented in the RPMDrate code.~\cite{RPMDrate} Its application to various gas phase chemical reactions has proven that the method is very accurate for estimating thermal rate coefficients even in the most challenging benchmark cases.~\cite{Suleimanov-etal:11,RPMD:MuH,Guo2012,RPMD:DMuH,RPMD2015,RPMDH5plus} 

Despite the instantaneous success of RPMDrate code,~\cite{WEB} the current version is restricted to a limited number of chemical reactions for which the underlying potential energy surfaces (PESs) are available in an analytical form. For the code to become a generally useful tool, efficient ways to couple RPMD with electronic structure evaluations are required. In principle, PES can be calculated “on-the-fly” but even with the most advanced supercomputers it is extremely CPU-intensive and is generally limited to fairly short propagation times. This challenge has been partially solved by approximating a limited number of quantum-mechanical calculations (typically tens of thousands), constructing a PES using the permutation invariant polynomial-neural network (PIP-NN) method.~\cite{PIPNN1,PIPNN2,PIPNN3} However, during preliminary RPMDrate simulations for several polyatomic systems, convergence issues have been detected due to artifacts in the PIP-NN PESs resulted from a lack of points in data sets in certain areas (see, e.g., the Supporting Information file of Ref.~\cite{PIPNN4}). As compared to classical trajectories which are normally used for verification of the PESs, RPMD trajectories provide more enhanced sampling of the phase space by ring polymer beads which could enter the potential artifact zones.

During the last years, the application of machine learning to constructing PESs has gained a lot of attention.~\cite{Behler,ArtrithKolpak2015NNP,Behler2011NNP,0953-8984-26-18-183001,Boes2016NNP-ReaxFF-comparison,Dolgirev2016,Gastegger2015high,Manzhos2015neural,NatarajanMorawietzBehler2015water,Lubbers2018,Smith2017,Kolb2017,schutt2017schnet,Bartok,GAP2014,DeringerCsanyi2016carbon,Deringer2018Boron,Grisafi2018tensorial,Shapeev2016-MTP,Thompson2015316,BotuRamprasad2015-MLIP,LiKermodeDevita2015lotf,Kruglov2017energy-free,sGDML,zhang2018-NNwater,ryczko2018-NNgraphene,kanamori2018-ML-PES-NEB,yao2018-NNtensormol} The methods are based on neural networks,~\cite{Behler,ArtrithKolpak2015NNP,Behler2011NNP,0953-8984-26-18-183001,Boes2016NNP-ReaxFF-comparison,Dolgirev2016,Gastegger2015high,Manzhos2015neural,NatarajanMorawietzBehler2015water,Lubbers2018,Smith2017,Kolb2017,schutt2017schnet,zhang2018-NNwater,ryczko2018-NNgraphene,yao2018-NNtensormol}
Gaussian processes,~\cite{Bartok,GAP2014,DeringerCsanyi2016carbon,Deringer2018Boron,Grisafi2018tensorial}
and other methods.~\cite{Shapeev2016-MTP,Thompson2015316}
Also closely related are energy-free ({\it i.e.}, non-conservative) machine-learning force fields.~\cite{BotuRamprasad2015-MLIP,LiKermodeDevita2015lotf,Kruglov2017energy-free}
Among those is the Moment Tensor Potential (MTP).~\cite{Shapeev2016-MTP,Gubaev2018-active-mlip}
We use MTP as the interatomic interaction model in this work.

The goal of this work is to propose an algorithm of automatically constructing an approximation to the reference PES for any given molecular system for subsequent calculation of RPMD thermal rate coefficients. The main challenge in automatically constructing such an approximation is to automatically assemble the training set that can be used to fit a good potential. A natural idea would be to use RPMDrate itself to sample the needed configurations for training, but the original version of RPMDrate requires a fitted potential to run. This seems to be a vicious circle: we need a training set in order to fit a potential, while we need a potential in order to sample a relevant training set. We resolve this challenge by applying the active learning (AL) approach, proposed in Ref.~\cite{Podryabinkin} for linearly parametrized potentials and extended to nonlinearly parametrized models in Refs.~\cite{Gubaev,Gubaev2018-active-mlip}. The idea of the approach is to let RPMDrate sample the needed configurations, and for each configuration decide on-the-fly whether a potential can yield reliable energies and forces or it needs to be trained on this configuration. The underlying algorithm for choosing configurations for training is based on a D-optimality criterion for selecting the configurations in the training set (after computing its energy and forces using an {\it ab initio} potential). The core of this criterion is the so-called maxvol algorithm, proposed in Ref.~\cite{Zamarashkin}. We refer to the combined approach as AL-MTP (active-learning moment tensor potential).

In this paper we propose and test a combination of AL-MTP and RPMDrate for predicting chemical reaction thermal rate coefficients. For the present study, we have selected two exemplifying systems, namely, OH + H$_2$ $\to$ H + H$_2$O and CH$_4$ + CN $\to$ CH$_3$ + HCN, recently studied using RPMDrate~\cite{Suleimanov2017A, Suleimanov2017B}. The RPMD rate coefficients and the corresponding analytical PESs~\cite{Chen,Suleimanov2017B} for these chemical reactions were readily available to us at the time we started this project. As our main purpose is to demonstrate the feasibility of our new approach, we consider these PESs as {\it ab initio} models and compare the rate coefficients predicted by these models to the ones calculated using the MTPs. We emphasize that although the practical purpose would be to fit machine-learning PESs to accurate quantum-mechanical models and hence calculate accurate reaction rates, the purpose of this work is to test the accuracy of our approach and hence we fit our PESs to the existing accurate and efficient PESs for which can compute the reaction rates used as a reference for our models.

\section{Methodology}
\subsection{Machine-Learning Interatomic Potential} \label{MLIP}

\subsubsection{Moment Tensor Potentials} \label{MTPdescription}

We assume that the energy of a configuration is partitioned into a sum of contributions of each of the $n$ atoms $E = \sum_{i=1}^n V_i$.
Each contribution is further expanded as a linear combination of basis functions $B_{\alpha}$,
\begin{align} 
\label{MTPrepresentation}
\displaystyle 
V_i &= \sum \limits_{\alpha} \xi_{\alpha} B_{\alpha},
\end{align}
where $\xi_\alpha$ are the parameters of the potential that are found (regressed) from the data.
The basis functions $B_{\alpha}$ depend on the atomic environment of the $i$-th atom consisting of all $j$-th atoms  that are within the distance of $R_{\rm cut}$ from the $i$-th atom.
The environment is expressed by the interatomic vectors $r_{ij}$ and the types of atoms $z_i$ and $z_j$. 
In order to account for all the physical symmetries, we introduce the moment tensor descriptors \cite{Shapeev2016-MTP}
\begin{equation}\label{Moments}
M_{\mu,\nu}(\bm r_i)=\sum_{j} f_{\mu}(|r_{ij}|,z_i,z_j) \underbrace {r_{ij}\otimes...\otimes r_{ij}}_\text{$\nu$ times}.
\end{equation}
Here the symbol ``$\otimes$'' denotes the outer product (so that $r_{ij}\otimes r_{ij}$ is a matrix, $r_{ij}\otimes r_{ij}\otimes r_{ij}$ is a three-dimensional tensor, etc.).
The first part, $f_{\mu}(|r_{ij}|,z_i,z_j)$, can be thought of as the radial part of the descriptors, while  $r_{ij}\otimes...\otimes r_{ij}$ is the angular part.
The radial part is further expanded as
\begin{align} \label{RadialPart}
\displaystyle
f_{\mu}(|r_{ij}|,z_i,z_j) = \sum_{\beta} c^{(\beta)}_{\mu, z_i, z_j} \varphi_\beta (|r_{ij}|),
\end{align}
where $c^{(\beta)}_{\mu, z_i, z_j}$ is another set of parameters to be fitted and $\varphi_\beta$ are the radial basic functions (expressed through the Chebyshev polynomials and ensuring a smooth cut-off to $0$ for $r>R_{\rm cut}$).
One can think of the functions $f_{\mu}$ as the ones that define the shells of neighboring atoms, while the coefficients $c^{(\beta)}_{\mu, z_i, z_j}$ express the relative weights of atomic species $z_j$ in the $\mu$-th shell of the $i$-th atom.

We then construct our basis functions $B_{\alpha}$ as different contractions of the moment tensor descriptors \eqref{Moments} to a scalar, such as
\begin{align*}
B_0(\bm r_i) &= M_{0,0}(\bm r_i),
\\
B_1(\bm r_i) &= M_{0,0}(\bm r_i) M_{1,0}(\bm r_i),
\\
B_2(\bm r_i) &=M_{0,2}(\bm r_i):M_{1,2}(\bm r_i),...
\end{align*}

\noindent We denote the parameters of MTP to be fitted by ${\bm {\theta}} := (\xi_{\alpha}, c^{(\beta)}_{\mu, z_i, z_j})$ and hence we denote the MTP energy of a configuration $\bm x$ by
$E = E({\bm {\theta}}; \bm x)$.

\subsubsection{Fitting} \label{Fittingdescription}

Let $\{\bm x^{(k)} \}$ be a training set with $K$ configurations.
Each configuration is supplied with an {\it ab initio} energy $E^{\rm AI}(\bm x^{(k)})$ and forces $f^{\rm AI}_i(\bm x^{(k)})$ on each of the atoms.
The fitting consists of finding the parameters $\bm \theta$ that minimize the following loss function
\begin{equation} \label{Fitting}
\displaystyle
L({\bm {\theta}}) = \sum \limits_{k=1}^K \left[ \left(E^{\rm AI}(\bm x^{(k)}) - E({\bm {\theta}}; \bm x^{(k)}) \right)^2 + w_{\rm f} \sum_{i=1}^n \left| f^{\rm AI}_i(\bm x^{(k)}) - f_{i}({\bm {\theta}}; \bm x^{(k)}) \right|^2 \right] \to \operatorname{min},
\end{equation} 
where $w_{\rm f}$ is a non-negative weight expressing the importance of forces relative to energy in the fitting. 

\subsubsection{Active Learning} \label{ALdescription}

Within the active learning concept, we construct the training set adaptively.
To achieve that, we need an algorithm that will decide whether to include a given configuration ${\bm x}^*$ that is  generated by the RPMDrate code. To that end, we need a new concept---\emph{active set}.
Suppose that the number of parameters ${\bm\theta}$ is $m$.
The \emph{active set} is then a subset of size $m$ of the training set (for convenience denoted by $\bm x^{(1)},\ldots,\bm x^{(m)}$) that maximizes the determinant
\[
\left|\begin{matrix}
\frac{\partial E}{\partial \theta_1}\left( {\bm \theta}; \bm x^{(1)} \right) & \ldots & \frac{\partial E}{\partial \theta_m}\left({\bm \theta}; \bm x^{(1)}\right) \\
\vdots & \ddots & \vdots \\
\frac{\partial E}{\partial \theta_1}\left({\bm \theta}; \bm x^{(m)}\right) & \ldots & \frac{\partial E}{\partial \theta_m}\left({\bm \theta}; \bm x^{(m)}\right) \\
\end{matrix}\right|.
\]
In order to find the \emph{active set} we use the so-called maxvol algorithm proposed in Ref.~\cite{Zamarashkin}.
For a configuration $\bm x^*$, we then define its extrapolation grade $\gamma(\bm x^*)$ as the maximum, by the absolute value, factor by which the above determinant can increase if we try to replace each $\bm x^{(i)}$ by $\bm x^*$.
We emphasize that $\gamma(\bm x^*)$ does not depend on the {\it ab initio} data, it depends only on the geometric information of the configuration $\bm x^*$.
Thus, it is not necessary to carry out {\it ab initio} calculations to calculate the extrapolation grade. 

In order to formulate our active learning algorithm, we introduce two thresholds: $\gamma_{\rm th}$ and $\Gamma_{\rm th}$, $1 < \gamma_{\rm th} < \Gamma_{\rm th}$.
These thresholds define the bounds of permissible extrapolation.
Thus our AL algorithm can be systemized as follows: 

\begin{itemize}
\item For each configuration $\bm x^*$ occurring in the RPMDrate simulation, we calculate $\gamma(\bm x^*)$. If $\gamma(\bm x^*) < \gamma_{\rm th}$ then $\bm x^*$ will not be added to the training set. Otherwise, there are two possibilities:

\begin{itemize}

\item[a.] $\gamma_{\rm th} \leq \gamma(\bm x^*) < \Gamma_{\rm th}$.
	In this case, we think of $\gamma(\bm x^*)$ as sufficiently high for $\bm x^*$ to be added to the training set, but not too high to terminate the RPMDrate simulation.
	Hence, in this case, we mark (save to a file) the configuration $\bm x^*$ and proceed with the RPMDrate simulation.
\item[b.] $\gamma(\bm x^*) \geq \Gamma_{\rm th}$.
In this case, the extrapolation grade is too high, therefore we add $\bm x^*$ to the training set and terminate the RPMDrate simulation.
We then update the \emph{active set} with the marked configurations, calculate their {\it ab initio} energies and forces, add them to the training set, refit the potential, and repeat the entire RPMDrate simulation from the beginning.
\end{itemize}
\end{itemize}
As a result, our algorithm will restart RPMDrate several times until the training set covers well the needed region in the phase space.
We emphasize that the potential is fixed at each RPMDrate run, thus ensuring that the code samples a proper canonical ensemble at each run.

Through the algorithm described above, our potential is trained in a fully automatic manner, lifting the need in tedious manual analysis of the quality of the PES being constructed.
The scheme of our AL-MTP algorithm is shown in Fig.~\ref{fig:AL}.

\begin{figure} \begin{center}
\includegraphics[width=6.0in, height=3.5in, keepaspectratio=false]{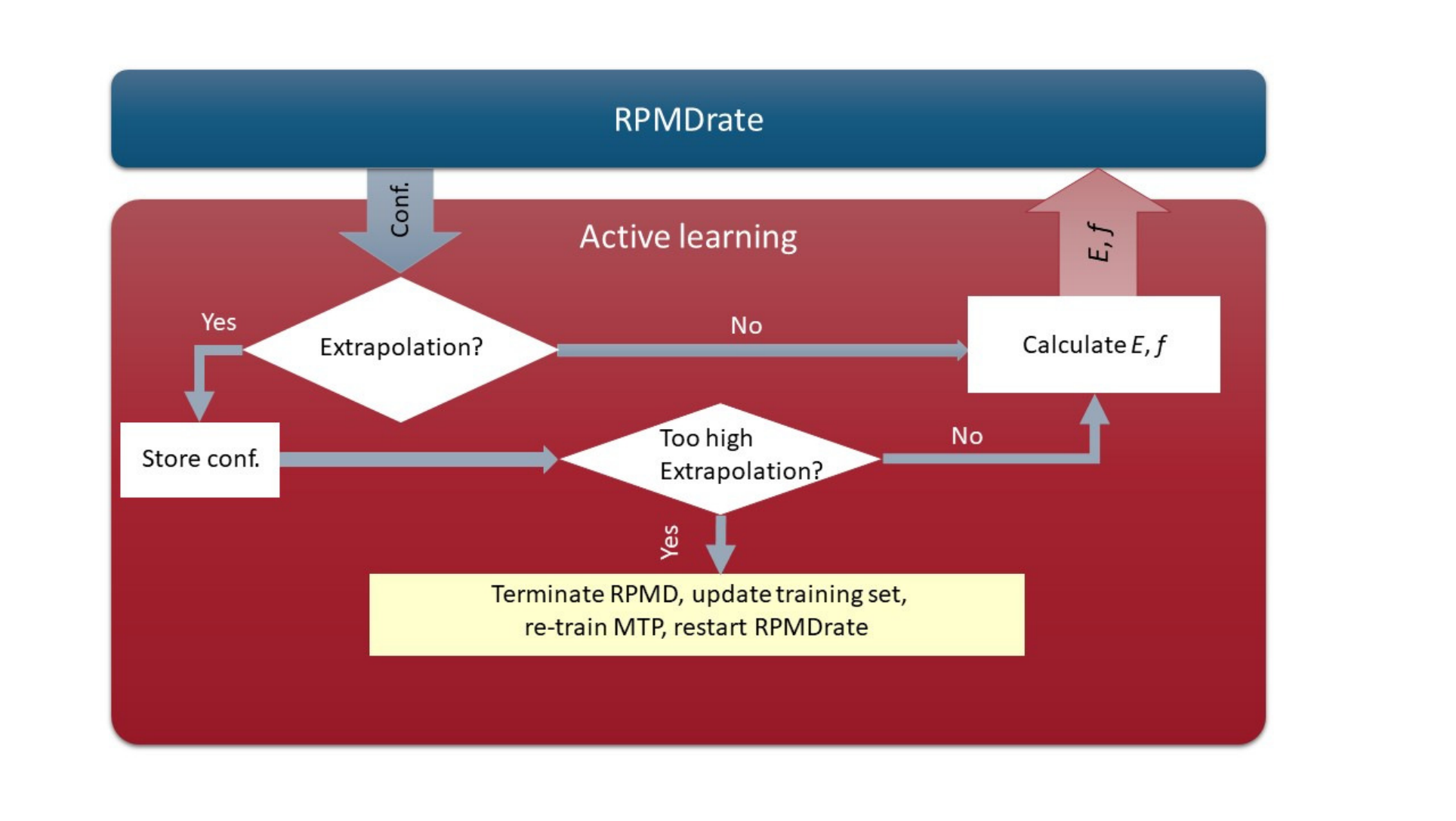}
\caption{\label{fig:AL} Active learning scheme. The RPMDrate code generates a configuration for which we calculate an extrapolation grade. If the grade is low,  we calculate the energy and forces for this configuration and continue the RPMDrate simulation. Otherwise, if the grade is high, but not too high to terminate the RPMDrate run,  we mark (save to a file) the configuration and proceed with the RPMDrate run. Finally, if the extrapolation grade is too high,  we terminate RPMDrate, update the training set, re-train the potential and restart the entire RPMDrate calculation.}
\end{center} \end{figure} 

\subsection{Application to OH + H$_2$ $\to$ H + H$_2$O and CH$_4$ + CN $\to$ CH$_3$ + HCN} 

We apply our AL algorithm in combination with MTPs to the calculation of RPMD rate coefficients for the following two representative chemical reactions: OH + H$_2$ $\to$ H + H$_2$O and CH$_4$ + CN $\to$ CH$_3$ + HCN. Below we show that the AL-MTP algorithm is capable of accurate prediction of chemical reaction rate coefficients for various  temperatures and different number of ring polymer beads for both systems.

\subsubsection{RPMDrate computational details}
We carry out the RPMD computations using the RPMDrate code which is well-documented in Ref.~\cite{RPMDrate}. Below we only briefly describe the key steps of the RPMDrate computational procedure. 
The rate coefficient is calculated using the Bennett-Chandler factorization~\cite{chandler1978,doi:10.1021/bk-1977-0046.ch004} as a product of a static (centroid density quantum transition state theory (QTST) rate coefficient, $k_{\rm QTST}$ ) and a dynamic (ring polymer transmission coefficient, $\kappa $) factors. 
The first step is the construction of potential of mean force (PMF) $W(\xi)$ along the dimensionless reaction coordinate $\xi$ defined in terms of two dividing surfaces given by Eqs.(4-10) in Ref.~\cite{RPMDrate}. The profile connects the reactant ($\xi = 0$) and transition state ($\xi = 1$) regions. We generate this profile using the umbrella integration technique~\cite{Kastner2005, Kastner2006} and use to calculate $k_{\rm QTST}$ . The second step is the calculation of $\kappa$ using a combination of constrained (parent) and unconstrained (child) trajectories. We perform steps consequently in order to detect the maximum value of $W(\xi^{\#})$ during the first step and to start the calculation of $\kappa$ from the coordinate $\xi^{\#}$ (for parent trajectory) during the second step. The final rate coefficient is given by the product of two factors, $k_{\rm RPMD} = k_{\rm QTST} \times \kappa$.

We study the first reaction, OH + H$_2$ $\to$ H + H$_2$O, at $T = 300 ~{\rm K}$ and $T = 1000 ~{\rm K}$ with $n_{\rm beads} = 1$ at both temperatures, $n_{\rm beads} = 128$ at the low temperature and $n_{\rm beads} = 16$ at the high temperature. We run the second reaction, CH$_4$ + CN $\to$ CH$_3$ + HCN, at $T = 300 ~{\rm K}$ and $T = 600 ~{\rm K}$ with the same number of ring polymer beads at the low and the high temperatures as for the first reaction. 

The remaining input parameters for the RPMDrate simulations are similar to those used in numerous studies of thermally activated chemical reactions.~\cite{Suleimanov-etal:16}   
In order to obtain the PMF profiles for both chemical reactions, we divide the interval $-0.05 \le \xi \le 1.05$ into 111 windows of width 0.01. Umbrella force constant was set to $k_i = 2.72 ((T/{\rm K}) ~{\rm eV})$ for each window centered at $\xi_i$, $i = 1, \ldots, 111$. In every window, we run 80 constrained RPMD trajectories with the sampling period of 50 ps and the equilibration period of 15 ps. Finally, the propagation time step was set equal to 0.0001 ps. 

For the calculation of $\kappa $, we choose slightly different parameters depending on the chemical reaction. For the OH + H$_2$ system, all the calculations (except the computation with $n_{\rm beads} = 128$) are carried out at 20000 unconstrained child trajectories ($N_{\rm totalchild}$) with the equilibration time of 10 ps ($t_{\rm equilibration}$) and 100 child trajectories per one initially constrained configuration ($N_{\rm child}$). All the unconstrained child trajectories run for $t_{\rm child} = 0.05$ ps with the time step $dt = 0.00005$ ps. For the case of $n_{\rm beads} = 128$, we increase the number of the unconstrained child trajectories up to 25000 and the time step is set to 0.0001 ps. For CH$_4$ + CN system, we take the following parameters: $N_{\rm totalchild} = 50000$, $t_{\rm equilibration} = 5$ ps, $N_{\rm child} = 100$, $t_{\rm child} = 0.06$ ps, and $dt = 0.0001$ ps.

As it was mentioned above, we consider the potentials described in Refs.~\cite{Chen,Suleimanov2017B}  as {\it ab initio} models for the present exemplifying study. The potential for the OH + H$_2$ $\to$ H + H$_2$O reaction has been developed using the Neural Networks (NN) fitting~\cite{Chen} and is denoted as NN1 PES. Another potential, applied for the CH$_4$ + CN system, is a combination of various semi-empirical potentials~\cite{Suleimanov2017B} including 34 parameters that were obtained after fitting this potential on the dataset, describing the stationary points, the reaction path and the reaction swath. For simplicity, we shall call this potential CH$_4$+CN PES though we note that its original abbreviation is different (PES2017). We also note that the previous RPMD studies using these PESs demonstrated very good agreement with the experimental measurements of rate coefficients~\cite{Suleimanov2017A,Suleimanov2017B} . 

We fit MTP with 92 basis functions $B_{\alpha}$, 4 radial functions $f_\mu$ and 12 radial basis functions $\varphi_{\beta}$.
This results into approximately 300 and 500 MTP parameters for OH + H$_2$ and  CH$_4$ + CN, respectively.
We choose $R_{\rm cut} $ = 4  and 6~$\angstrom$, respectively, for these systems.
The active learning was performed with $\gamma_{\rm th} = 2$ and $\Gamma_{\rm th} = 10$, thus the interval of high but permissible \emph{grades} is $[2, 10)$.

As described above, we need to compute $k_{\rm QTST}$ (the first step of RPMDrate) and $\kappa$ (the second step of RPMDrate).
In order to obtain $k_{\rm QTST}$, we focus only on the region connecting the reactants with the transition state ({\it i.e.}, $\xi \in (-0.05, 1.05)$). During the second step -- computation of $\kappa$ -- the RPMD trajectories visit the products region, {\it i.e.}, $\xi > 1.05$. The geometries of the configurations in the reactants and products regions are different and, thus, we use a slightly different MTPs for the calculation of $k_{\rm QTST}$ and $\kappa$ trained on two datsets. More precisely, during the first RPMDrate step we form the reactants set ($k_{\rm QTST}$ set) that consists of configurations selected from the reactants region and learn on-the-fly the first MTP.
During the second RPMDrate step, we start from the MTP and the training set derived after the first step, update the training set with the additional configurations (the products set, or, $\kappa$ set) and learn on-the-fly the second MTP.
Thus, the two MTPs differ by their training sets---the first training set is a subset of the second one.
Having computed $k_{\rm QTST}$ and $\kappa$, respectively, by these two MTPs, we obtain the final RPMD rate coefficient. 

\section{Results and discussion}
 
The PMF profiles $W(\xi)$ for the OH + H$_2$ and  CH$_4$ + CN reactions are plotted in Figs.~\ref{fig:PMFsOH+H2} and~\ref{fig:PMFsCH4+CN}, respectively. For both reactions and two representative temperatures, the results obtained using MTPs are close to the {\it ab initio} profiles, the difference is less than 0.3 kcal/mol.
Time-dependent $\kappa $'s obtained by the MTP and {\it ab initio} models for the OH + H$_2$ and CH$_4$  + CN reactions are shown in Figs.~\ref{fig:TCsOH+H2}  and~\ref{fig:TCsCH4+CN}, respectively.
Similarly to the PMF profiles, the results obtained using MTP are in a very good agreement with the {\it ab initio} counterparts.
The values of the centroid density TST rate coefficient $k_{\rm QTST}$, the ring polymer recrossing factor $\kappa$ and the RPMD rate coefficient $k_{\rm RPMD}$ are also summarized for the OH + H$_2$ and CH$_4$ + CN reactions in Tables~\ref{tabl:CoeffsOH+H2} and~\ref{tabl:CoeffsCH4+CN}, respectively.
The agreement with the previous RPMD rate coefficients is very good, the relative root-mean-square deviation between MTP and the reference rate coefficients is about 20\% or less and is comparable to typical convergence error of the RPMDrate computational procedure~\cite{RPMDrate,Suleimanov-etal:16}.

The number of configurations selected in the reactants region ($k_{\rm QTST}$ set size), the products region ($\kappa$ set size) and the total training set sizes ($k_{\rm RPMD}$ set size) are reported in Table~\ref{tabl:TSsize}. As it could be seen, we select many more configurations from the reactants region than those we add from the products region (see Fig.~\ref{fig:histogram_configurations}).
The reason of it is as follows. During the first RPMDrate step, we need to approximate the PMF difference between the reactants and the transition state as accurate as possible due to its exponential contribution to $k_{\rm QTST}$. Thus, we need to predict the PMF profile across each umbrella window, especially near the transition state. This fact is illustrated in Fig.~\ref{fig:histogram_xi}. Indeed, most of configurations for both systems were selected for $\xi \in (0.95, 1.05)$, {\it i.e.}, near the transition state. 
Then, it happens that for the purposes of calculating RPMD rate coefficients a potential that is well-trained in the reactants region needs much less data to be fitted in the products region (only a few bonds significantly differ, while most of the bonds in the molecular systems are the same in both regions).

The size of a total training set significantly depends on the number of different atomic types in the molecule.
Note that the number of parameters $c^{(\beta)}_{\mu, z_i, z_j}$ grows as the square of number of atomic types (as they depend on pairs of types of interacting atoms, $(z_i,z_j)$).
Thus, there are 2.25 times more coefficients in the potential for the CH$_4$ + CN system than for the OH + H$_2$ one.
As it can be seen from Table \ref{tabl:TSsize}, we need approximately 2.25 times more configurations in the training sets for the CH$_4$ + CN system than for the OH + H$_2$ one.
This confirms that the number of coefficients grows quadratically with the number of atomic types and thus the proposed algorithms should be applicable for large molecular systems, direct description of which is problematic due to high dimensionality. 

	We additionally test how the accuracy improves when the number of MTP parameters increases.
	The test is done for the CH$_4$ + CN system, T = 300 K, $n_{\rm beads} = 1$.
	On Fig. \ref{fig:TSsize_vs_accuracy} the PMF profile and the transmission coefficient are plotted for three potentials, with 150, 250, and 500 parameters, respectively.
	As it could be seen, the training set size increases with the number of parameters, and so is the accuracy.

The remaining two factors that affect the size of our training set are the number of ring polymer beads and the temperature. Increasing the number of ring polymer beads leads to more enhanced phase space exploration thus more configurations are necessary in the training set. The same is valid for the temperature factor: as the temperature increases, the energy dispersion increases and therefore we need more configurations in the dataset in order to describe all possible energy levels.
We attribute both correlations to the fact that higher temperature and higher number of beads imply that we need to sample a larger region in the phase space and therefore collect more configurations for training.
In any case, the maximal training set size is less than 5000, thus, we needed to carry out less than 5000 {\it ab initio} calculations in order to obtain an MTP for both exemplifying chemical systems considered in the present study.

\section{Conclusions}
In summary, we propose a fully automated procedure for calculating ring polymer molecular dynamics (RPMD) rate coefficients using the potential energy surface (PES) generated on-the-fly by the moment tensor potentials (MTP) with active learning (AL). The procedure follows the original Bennett-Chandler factorization implemented in the RPMDrate code which splits the calculation in two steps---a static (centroid density quantum transition state theory (QTST) rate coefficient) and a dynamic (ring polymer transmission coefficient) factors.
During each step, the active-learning algorithm accumulates automatically the dataset sample, ensuring that the fit of the PES is appropriate for calculating the RPMD rate coefficient for a given temperature and number of ring polymer beads. 
In order to determine whether the current point should be added to the training or not, set we calculate the energy gradient with respect to the parameters of the potential and the so-called extrapolation grade. If the extrapolation grade is greater than the lower bound of permissible extrapolation, we mark the current point (save to a file). If the extrapolation grade is greater than the upper bound of permissible extrapolation,  we terminate RPMDrate, update the training set using maxvol algorithm and refit the potential. Such an approach ensures that the final machine-learning PES model avoids artifacts during exploration of the phase space by RPMD trajectories which  have been observed for several PESs fitted by neural networks~\cite{PIPNN4}. 
The methodology is tested on two representative thermally activated chemical reactions, namely, OH + H$_2$ and CH$_4$ + CN which were previously studied by RPMD.~\cite{Suleimanov2017A,Suleimanov2017B} The deviation of the present  RPMD rate coefficients  obtained using the AL-MTP approach from the reference values is within the convergence error of the RPMDrate  computational procedure. 

In future, we plan to extend our methodology to chemical reactions which proceed via complex formation in order to propose a completely general tool for calculating RPMD rate coefficients for any polyatomic chemical reactions. 
In principle, the Bennett-Chandler factorization can be also implemented in this case~\cite{Suleimanov-etal:16}  though the contribution from the real-time propagation of the dynamic factor significantly increases leading to possible alterations to the AL-MTP algorithm. This work is currently on-going. 
Finally, we would like to note that our AL-MTP approach could be used in calculations of other dynamical properties  (such as RPMD diffusion coefficients), applicability of the algorithm does not depend on a physical quantity predicted.

\section{Acknowledgements}
The work of I.S.N.\ and A.V.S.\ was supported by the Russian Science Foundation (grant number 18-13-00479).
Y.V.S.\ thanks the European Regional Development Fund and the Republic of Cyprus for support
through the Research Promotion Foundation (Project Cy-Tera NEA ${\rm Y\Pi O\Delta OMH}$/ ${\rm \Sigma TPATH}$/0308/31). We thank Jesus Castillo for providing analytical gradients for the NN1 PES. 
We also thank our colleagues, Konstantin Gubaev and Evgeny Podryabinkin, for giving advance access to the code implementing AL-MTP.
This work was performed, in part, by A.V.S. at the Center for Integrated Nanotechnologies, an
Office of Science User Facility operated for the U.S. Department of Energy (DOE) Office of Science by
Los Alamos National Laboratory (Contract DE-AC52-06NA25396) and Sandia National Laboratories
(Contract DE-NA-0003525).

\bibliographystyle{pccp}
\bibliography{article}

\newpage
\begin{figure}[h!] \begin{center}
\begin{turn}{270}
\includegraphics[trim = 0cm 0cm 10.5cm 0cm,clip=true, scale=0.6, keepaspectratio=false]{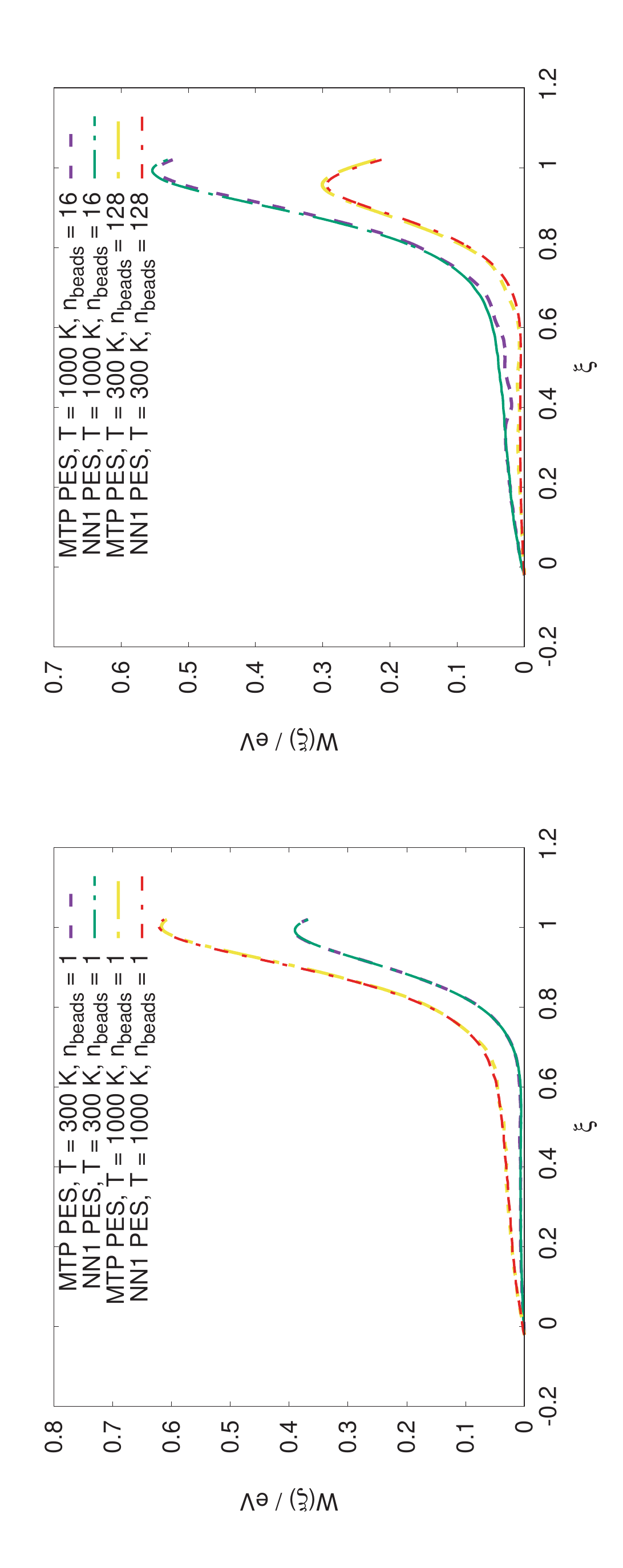}
\end{turn}
\caption{\label{fig:PMFsOH+H2} Comparison of potential of mean force profiles for the OH + H$_2$ $\to$ H + H$_2$O reaction calculated by the Moment Tensor Potential (MTP) PES and NN1 (reference PES) under various temperatures and number of beads.}
\end{center} \end{figure}

\begin{figure}[h!] \begin{center}
\begin{turn}{270}
\includegraphics[trim = 0cm 0cm 10.5cm 0cm,clip=true, scale=0.6, keepaspectratio=false]{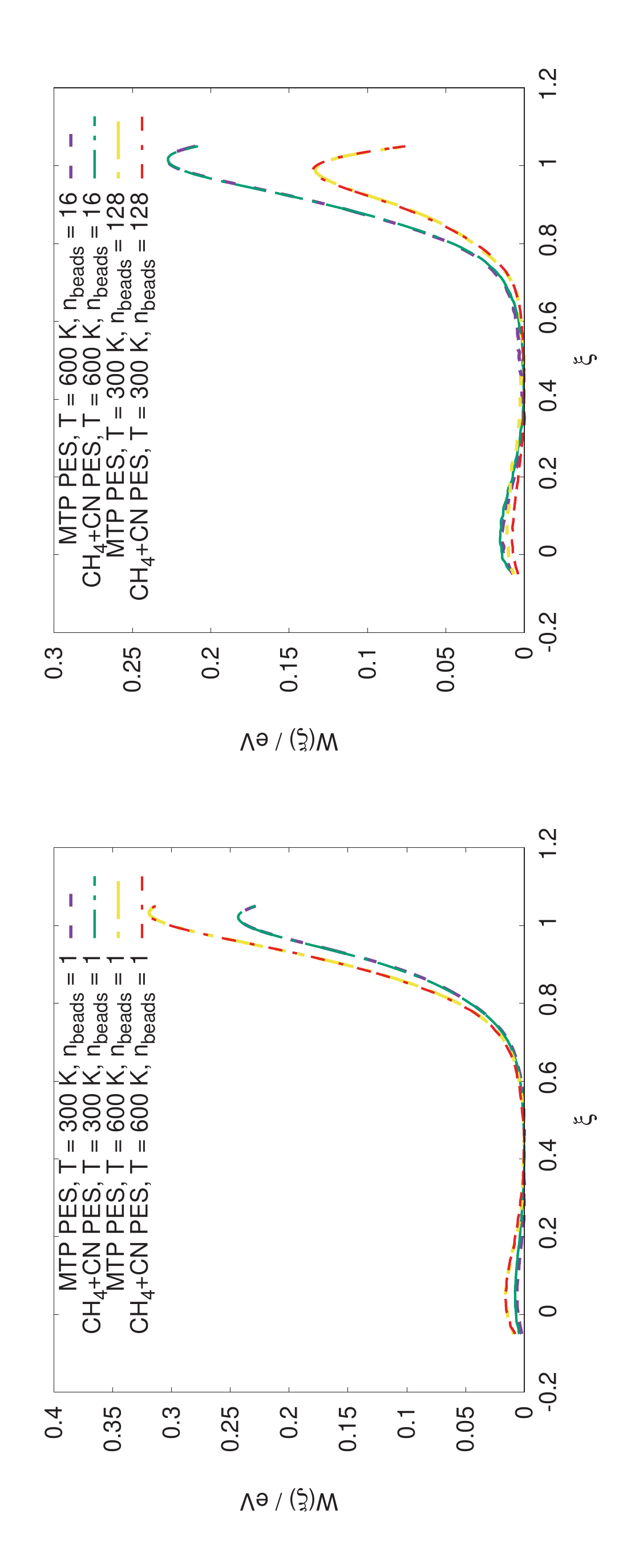}
\end{turn}
\caption{\label{fig:PMFsCH4+CN} Comparison of potential of mean force profiles for the CH$_4$ + CN $\to$ CH$_3$ + HCN reaction calculated by the Moment Tensor Potential (MTP) PES and CH$_4$ + CN (reference PES) under various temperatures and number of beads.}
\end{center} \end{figure}

\begin{figure}[h!] \begin{center}
\begin{turn}{270}
\includegraphics[trim = 0cm 0cm 10.5cm 0cm,clip=true, scale=0.6, keepaspectratio=false]{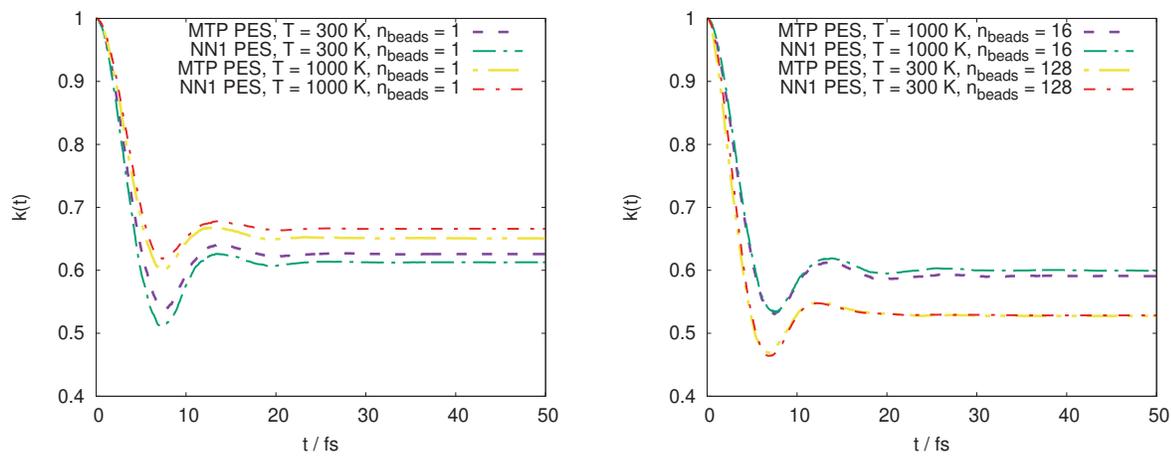}
\end{turn}
\caption{\label{fig:TCsOH+H2} Comparison of the time-dependent ring polymer transmission coefficients for the OH + H$_2$ $\to$ H + H$_2$O reaction calculated by the Moment Tensor Potential (MTP) PES and NN1 (reference PES) under various temperatures and number of beads.}
\end{center} \end{figure}

\begin{figure}[h!] \begin{center}
\begin{turn}{270}
\includegraphics[trim = 0cm 0cm 10.5cm 0cm,clip=true, scale=0.6, keepaspectratio=false]{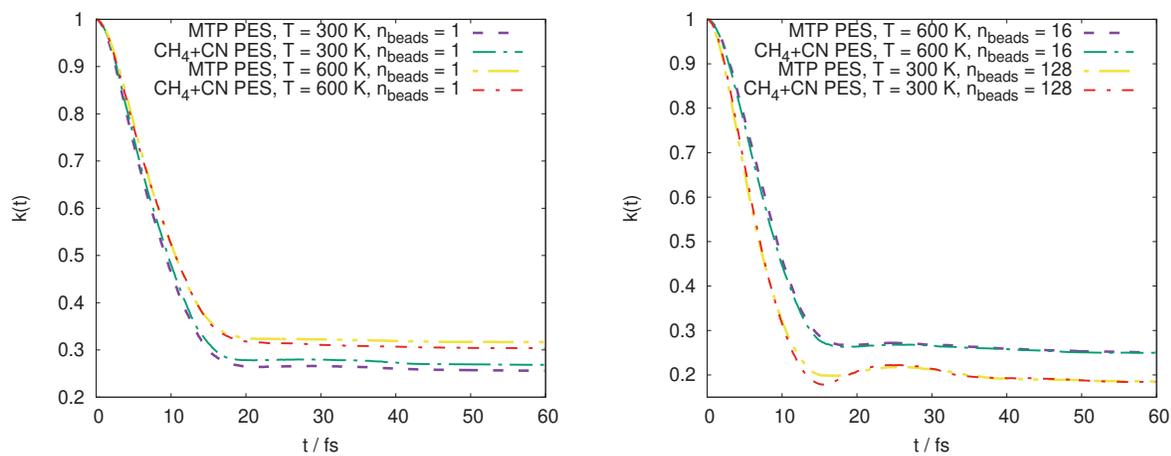}
\end{turn}
\caption{\label{fig:TCsCH4+CN} Comparison of the time-dependent transmission coefficients for the CH$_4$ + CN $\to$ CH$_3$ + HCN reaction obtained by Moment Tensor Potential (MTP) and CH$_4$ + CN (reference PES) under various temperatures and number of beads.}
\end{center} \end{figure}

\begin{table}[h!]
\begin{center}
\begin{tabular}{c|cccc} \hline \hline
 & T = 300 K & T = 300 K & T = 1000 K & T = 1000 K \\
 & $n_{\rm beads}$ = 1 & $n_{\rm beads}$ = 128 & $n_{\rm beads}$ = 1 & $n_{\rm beads}$ = 16 \\ \hline
$k_{\rm QTST}^{\rm AI}$ (cm$^3$ s$^{-1}$) & $5.74 \times 10^{-16}$ & $2.37 \times 10^{-14}$ & $2.78 \times 10^{-12}$ & $3.72 \times 10^{-12}$ \\ 
$k_{\rm QTST}^{\rm MTP}$ (cm$^3$ s$^{-1}$) & $5.37 \times 10^{-16}$ & $1.84 \times 10^{-14}$ & $2.91 \times 10^{-12}$ & $3.97 \times 10^{-12}$ \\ 
error (\%) & 6.5 \% & 22.3 \% & 4.7 \% & 6.7 \% \\ \hline
$\kappa^{\rm AI}$ & 0.613 & 0.528 & 0.666 & 0.599 \\ 
$\kappa^{\rm MTP}$ & 0.626 & 0.527 & 0.649 & 0.589 \\ 
error (\%) & 2.1 \% & 0.2 \% & 2.6 \% & 1.7 \% \\ \hline
$k_{\rm RPMD}^{\rm AI}$ (cm$^3$ s$^{-1}$) & $3.52 \times 10^{-16}$ & $1.25 \times 10^{-14}$ & $1.85 \times 10^{-12}$ & $2.23 \times 10^{-12}$ \\
$k_{\rm RPMD}^{\rm MTP}$ (cm$^3$ s$^{-1}$) & $3.36  \times 10^{-16}$ & $9.70 \times 10^{-15}$ & $1.89 \times 10^{-12}$ & $2.34 \times 10^{-12}$ \\
error (\%) & 4.5 \% & 22.4 \% & 2.2 \% & 4.9 \% \\ 
\hline
\hline
\end{tabular}
\caption{\label{tabl:CoeffsOH+H2} Comparison of the quantum transition state theory (QTST) rate coefficient $k_{\rm QTST}$, ring polymer transmission coefficient $\kappa$, and final rate coefficient $k_{\rm RPMD}$ calculated by the NN1 and MTP PESs for the OH + H$_2$ system under various conditions.} 
\end{center}
\end{table}

\begin{table}[h!]
\begin{center}
\begin{tabular}{c|cccc} \hline \hline  
 & T = 300 K & T = 300 K & T = 600 K & T = 600 K \\
 & $n_{\rm beads}$ = 1 & $n_{\rm beads}$ = 128 & $n_{\rm beads}$ = 1 & $n_{\rm beads}$ = 16 \\ \hline
$k_{\rm QTST}^{\rm AI}$ (cm$^3$ s$^{-1}$) & $1.69 \times 10^{-13}$ & $1.13 \times 10^{-11}$ & $6.10 \times 10^{-12}$ & $3.63 \times 10^{-11}$ \\ 
$k_{\rm QTST}^{\rm MTP}$ (cm$^3$ s$^{-1}$) & $1.61 \times 10^{-13}$ & $1.35 \times 10^{-11}$ & $6.17 \times 10^{-12}$ & $3.48 \times 10^{-11}$ \\ 
error (\%) & 4.7 \% & 19.5 \% & 1.1 \% & 4.1 \% \\ \hline
$\kappa^{\rm AI}$ & 0.267 & 0.184 & 0.304 & 0.250 \\ 
$\kappa^{\rm MTP}$ & 0.256 & 0.185 & 0.317 & 0.251 \\ 
error (\%) & 4.1 \% & 0.5 \% & 4.3 \% & 0.4 \% \\ \hline
$k_{\rm RPMD}^{\rm AI}$ (cm$^3$ s$^{-1}$) & $4.51 \times 10^{-14}$ & $2.08 \times 10^{-12}$ & $1.85 \times 10^{-12}$ & $9.07 \times 10^{-12}$ \\
$k_{\rm RPMD}^{\rm MTP}$ (cm$^3$ s$^{-1}$) & $4.12 \times 10^{-14}$ & $2.50 \times 10^{-12}$ & $1.95 \times 10^{-12}$ & $8.73 \times 10^{-12}$ \\
error (\%) & 8.6 \% & 20.2 \% & 5.4 \% & 3.7 \% \\ 
\hline
\hline 
\end{tabular}
\caption{\label{tabl:CoeffsCH4+CN} Comparison of the quantum transition state theory (QTST) rate coefficient $k_{\rm QTST}$, ring polymer transmission coefficient $\kappa$, and final rate coefficient $k_{\rm RPMD}$ calculated by the CH$_4$ + CN and MTP PESs for the CH$_4$ + CN system under various conditions.} 
\end{center}
\end{table}

\begin{table}[h!]
\begin{center}
\begin{tabular}{c|c|c|c|c|c|c|c} \hline \hline 
\multicolumn{4}{c|}{OH + H$_2$ $\to$ H + H$_2$O} & \multicolumn{4}{c}{CH$_4$ + CN $\to$ CH$_3$ + HCN} \\ \hline
T, $n_{\rm beads}$ & $k_{\rm QTST}$ & $\kappa$ & $k_{\rm RPMD}$ & T, $n_{\rm beads}$ & $k_{\rm QTST}$ & $\kappa$ & $k_{\rm RPMD}$ \\ 
 & set size & set size & set size & & set size & set size & set size \\ \hline
300 K, 1 & 1401 & 96 & 1497 & 300 K, 1 & 3348 & 581 & 3929 \\ 
300 K, 128 & 1816 & 44 & 1860 & 300 K, 128 & 4138 & 380 & 4518 \\ 
1000 K, 1 & 1784 & 123 & 1907 & 600 K, 1 & 3904 & 544 & 4448 \\ 
1000 K, 16 & 2014 & 83 & 2097 & 600 K, 16 & 4572 & 320 & 4892 \\ \hline \hline 
\end{tabular}
\caption{\label{tabl:TSsize} Number of configurations selected in the reactants region ($k_{\rm QTST}$ set size), in the products region ($\kappa$ set size), and the total training set size ($k_{\rm RPMD}$ set size) for the OH + H$_2$ and CH$_4$ + CN systems.} 
\end{center}
\end{table}

\begin{figure}[h!] \begin{center}
\begin{turn}{270}
\includegraphics[trim = 0cm 0cm 10.5cm 0cm,clip=true, scale=0.6, keepaspectratio=false]{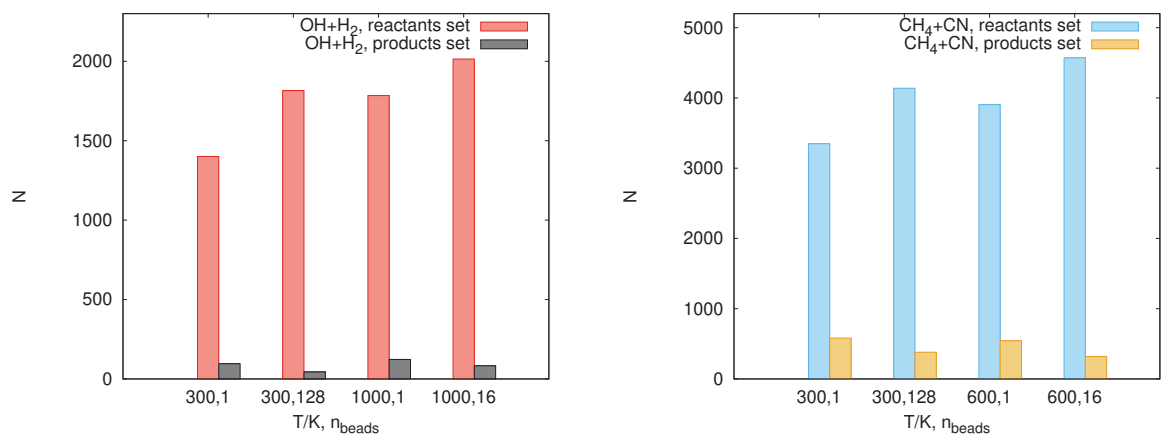}
\end{turn}
\caption{\label{fig:histogram_configurations} The reactants and products set sizes for the OH+H$_2$ and CH$_4$+CN systems. For both reactions under various conditions the largest number of configurations, $N$, was selected in the reactants region.}
\end{center} \end{figure}

\begin{figure}[h!] \begin{center}
\begin{turn}{180}
\includegraphics[trim = 0cm 0cm 0cm 3cm,clip=true, scale=0.8, keepaspectratio=false]{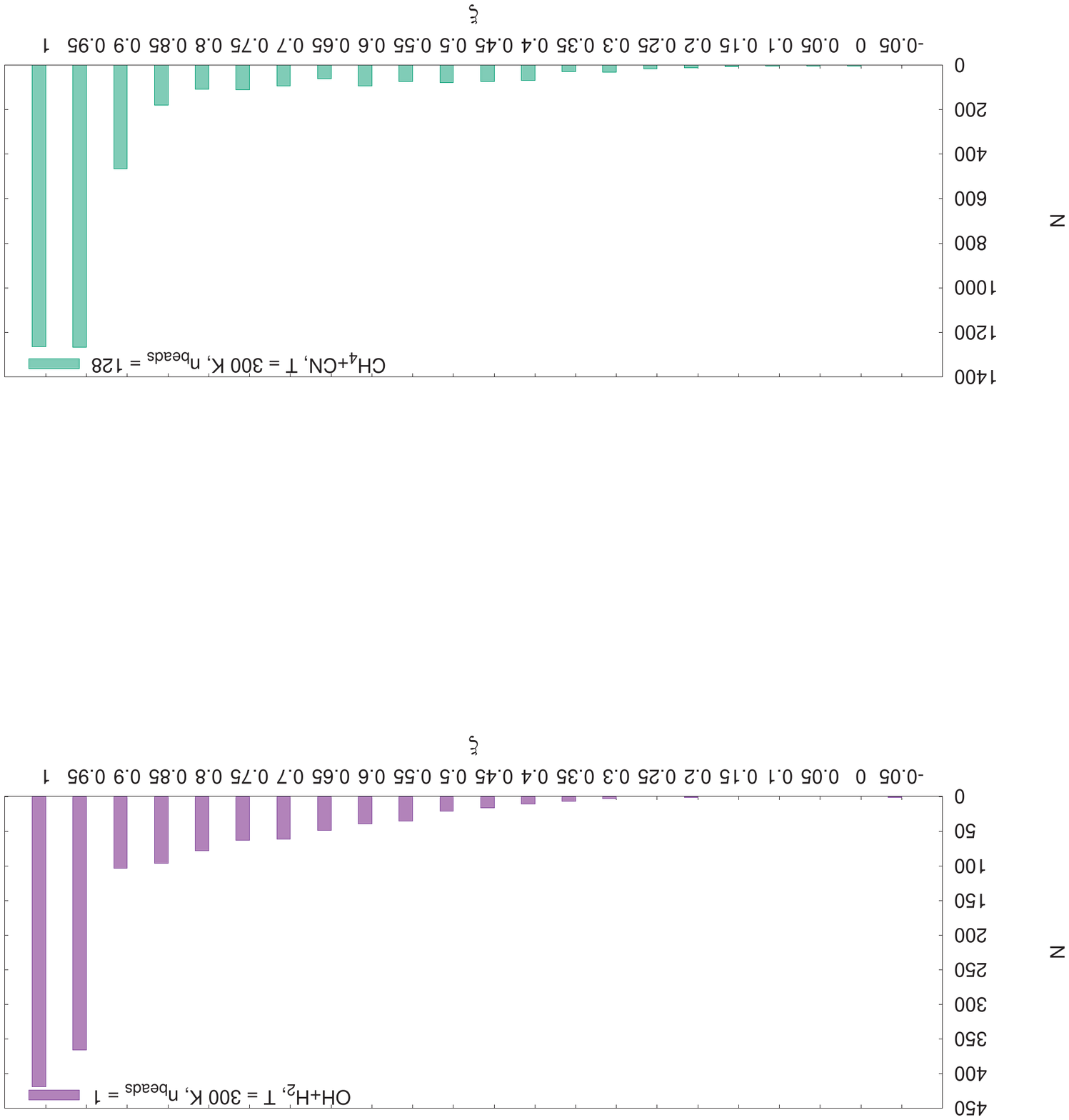}
\end{turn}
\caption{\label{fig:histogram_xi} Dependence of the number of configurations on the reaction coordinates for the OH+H$_2$ and CH$_4$+CN systems. The numbers are given for the intervals (-0.05, 0), (0, 0.05), \ldots, (1, 1.05). The transition state is located near the point $\xi = 1$, the largest number of configurations $N$ was selected around this point.}
\end{center} \end{figure}

\begin{figure}[h!] \begin{center}
\begin{turn}{270}
\includegraphics[trim = 0cm 0cm 5.5cm 0cm,clip=true, scale=0.65, keepaspectratio=false]{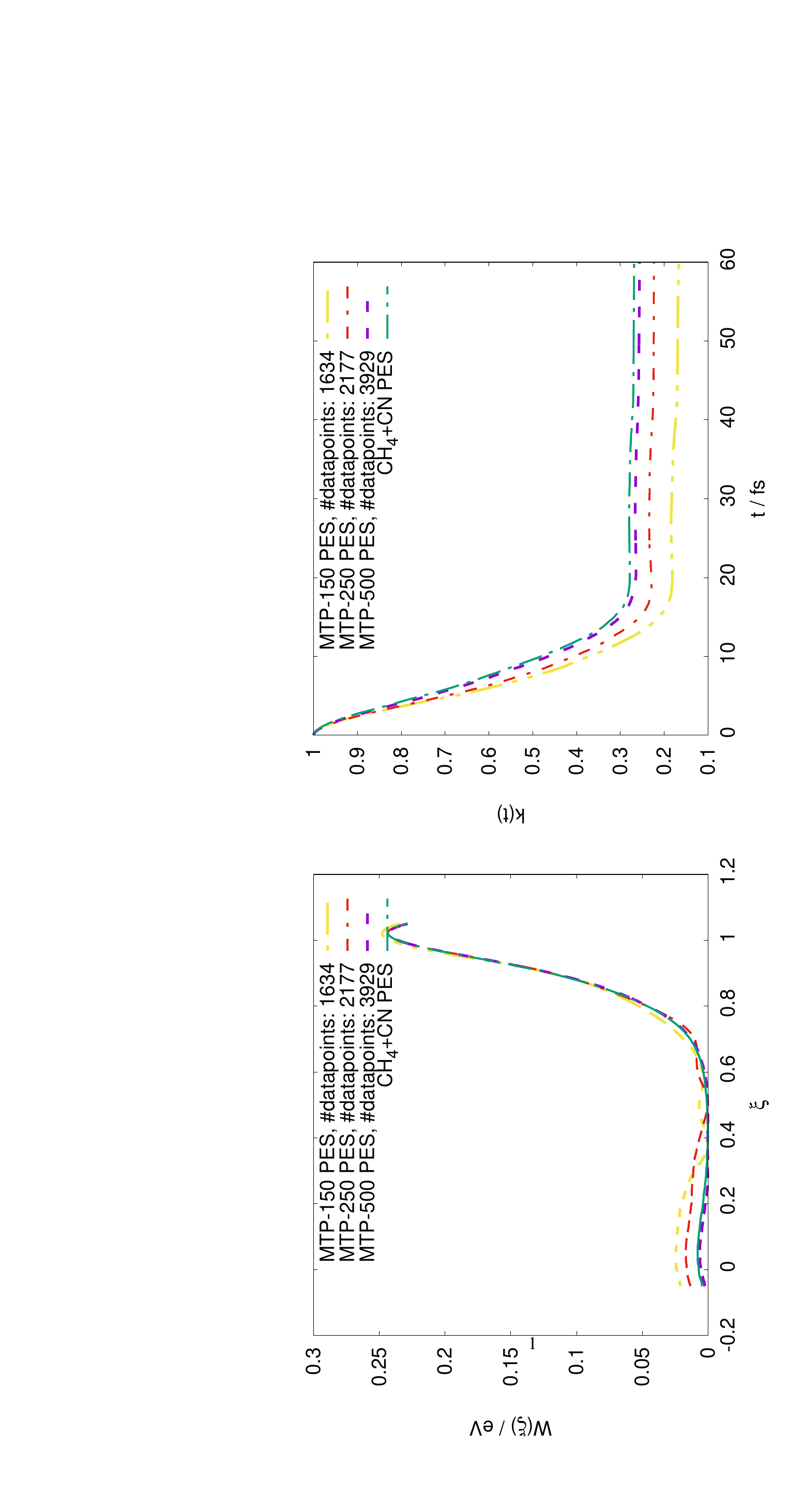}
\end{turn}
\caption{\label{fig:TSsize_vs_accuracy} The dependence of the accuracy of the potential of mean force and transmission coefficient on the number of parameters in Moment Tensor Potentials (150, 250, and 500) and on the dataset size.
		The potentials are labeled MTP-150, MTP-250, and MTP-500, respectively.
The data is for the CH$_4$ + CN system, T = 300 K, and $n_{beads} = 1$. The number of datapoints improves the accuracy of the calculated coefficients.}
\end{center} \end{figure}

\end{document}